# Influence of the structure defects on the magnetic properties of the FePt/Fe bilayer

**E.M. Plotnikova**[1,2], **I.I. Trushkin**[1,2], **D.A. Lenkevich**[3], **A.L.Kotelnikov**[3,4], **A.Cockburn**[5], **K.A. Zvezdin**[2,4]

[1] Moscow Institute of Physics and Technology State University, 9 Institutskiy lane, Dolgoprudny city, Moscow Region, 141700, Russian Federation
[2] Prokhorov General Physics Institute, RAS, 38 Vavilov Str., Moscow, 119991, Russian Federation
[3] Joint Institute for High Temperatures (JIHT), RAS, 13 Izhorskaya st. Bd.2, Moscow, 125412, Russian Federation
[4] Istituto P.M. srl, I-10141, via Grassi, 4, Torino, Italy
[5] Institute for Manufacturing Department of Engineering, University of Cambridge, 17 Charles Babbage Road, Cambridge, CB3 0FS, UK

E-mail: `ekaterina.plotnikova@phystech.edu`

**Abstract.** Thin magnetic multilayered films containing FePt have attracted a lot of attention recently due to their possible usage in ultra-high density magnetic storage. Although structure defects play a dramatic role in the magnetization process and influence magnetic properties in general this dependence haven't been studied thoroughly. The main aim of this work was to perform theoretical investigation of the magnetic properties of FePt and Fe/FePt thin films with high coercivity with respect to the structure defects such as anisotropy constant, magnetization saturation, exchange constant fluctuations and easy axis deviation. For selected defect patterns the coercive field dependence on layer thicknesses was analysed. Numerical study of the bilayer with hard magnetic layer having the planar anisotropy was carried on using micromagnetic calculations. Values of layers thickness have been found optimal for perspective applications, the dependence of the hysteresis loop shape upon the magnetization process has been shown and analysed.

## 1. Introduction

Thin film magnetism has become a highly active field of research during last decades. With the better understanding of the magnetic nanostructures physics follows a more device-oriented research effort. Recently new capabilities for such studies have been given, on the one hand, by the application of advanced lithography and pattern transfer [1], [2] techniques to the fabrication of nanoscale magnetic and by employing progressive measurement equipment; on the other hand, the spectacular development of calculation methods which has enabled a much deeper understanding of materials properties.

For pure bulk material such as Co or Ni values of magnetocristalline anisotropy (MCA) constant K usually do not exceed 1-2 $erg/cm^3$ (So, for fcc Co K = -1.2 $erg/cm^3$, for bcc Fe K = 0.481 $erg/cm^3$, for fcc Ni K = -0.056 $erg/cm^3$ [3], [4]). For some of the composed materials MCA constant was discovered to be significantly bigger, for instance, for CoPt K is around 5-7·$10^7$ $erg/cm^3$ [5]. Also, the chemically ordered $L1_0$ phase of FePt has shown large uniaxial MAE with the first-order anisotropy constant K up to 7·$10^7$ $erg/cm^3$. In combination with good intrinsic magnetic properties such as comparatively hight Curie temperature $T_C$ = 750 K, spontaneous magnetization at room temperature $J_S$ = 1.43 T [6], high ductility and good corrosion resistance the tetragonal FePt phase attracted much interest from both fundamental and application points of view. In the $L1_0$ phase the cubic symmetry is broken due to the stacking of alternate planes of the Fe atom and the Pt atom along the [001] direction [7]. It is well established that in this naturally layered ferromagnet the large MAE is mainly due to the contribution from the 5d element having large spin-orbit (s-o) coupling while the 3d element provides the exchange splitting of the 5d sub-lattice [8]. The increasing need for ultra-high density magnetic storage has caused the bit volume to shrink to the nanometer scale and numerous theoretical [9],[10],[11],[12],[13],[14] and experimental research [15],[16] have been done to demonstrate that a high degree of equilibrium $L1_0$ order is achievable even at the nanoscale if kinetic issues are overcome. Thus, FePt is a very promising material indeed.

Recently a plenty of theoretical studies have been done [17], [18], [19] and a number of experiments has been performed on FePt-based systems such as exchange-coupled $L1_0$ FePt/[Co/Ni]$_N$ [20], FePt/[Co/Pt] [21], FePt/Fe$_3$Pt [6] nanocomposite multilayers, FePt/CoCrNi [22], FePt/FeRh [23], FePt/CoFeTaB [24] bilayers, FePt thin films on Si [25] and CrV [26] substrate, FePt nanoparticles [27], [28], FePt/Fe thin films [29], [30], [31], [32], [33], [34]. Among them, FePt/Fe system has shown unique magnetic values. The big magnetocrystalline anisotropy constant of FePt and the large magnetic saturation polarization of Fe result in a significant remanence enhancement of the exchange-coupled hard/soft magnetic bilayers [29]. Although magnetic properties, particularly the magnetization reversal mechanism, and thermal stability of hard magnetic FePt alloy films with softer Fe layer has been studied extensively both experimentally and in the terms of micromagnetic approach [29], [30], [31], [32], [33],[35] there are no specific theoretical studies of defects influence on the FePt/Fe bilayer.

As it follows from the experiments [36], [37], [38] defects play a dramatic role in the magnetization process and influence magnetic properties in general. Furthermore, study of an exchange-coupled magnet (ECM) being carried out without defects employed would hardly show the correct results as far as in fact defects affect the hard magnetic layer much stronger than the soft one. Absence of the defects in the model leads to underestimating of the advantages of adding soft layer to the hard one. Therefore to get not only quantitatively but even qualitatively correct results it was highly desirable to consider defects in the framework of the micromagnetic model. However, experimental study of the role of defects seems to be a hard task since it is not possible to control the way defects appear during structure growth, deposition or annealing processes. Micromagnetic research gives a good possibility to study the role of defects as far as one can fix the set of defects for studied structure and thus becomes a very promising tool for such applications indeed.

## 2. Studied system

The system to be studied is an exchange-coupled bilayer being composed of hard $L1_0$-FePt and soft Fe magnetic layers as shown on Fig 1. Planar anisotropy is considered. The nonmagnetic layer which was added in order to simulate interlayer exchange correctly in the framework of our micromagnetic model has 0.5 nm thickness thereby performing 4% of whole structure.

The micromagnetic study of the reversal processes were performed by numerical integration of the Landau-Lifshitz-Gilbert equation using our micromagnetic code SpinPM based on the forth order Runge-Kutta method with an adaptive time-step control for the time integration and a mesh size 2,5×2,5 nm$^2$.

The magnetization parameters of hard magnetic layer are: the magnetization $M_S = 1139\ emu/cm^3$, the exchange energy $A = 1\times10^{-6} erg/cm$, the anisotropy constant $K_1 = 6 \times 10^7 erg/cm^3$. The magnetization parameters of soft magnetic layer are: the magnetization $M_S = 1671\ emu/cm^3$, the exchange energy $A = 2.8 \times 10^{-6} erg/cm$, the anisotropy constant $K_1 = 0\ erg/cm^3$.

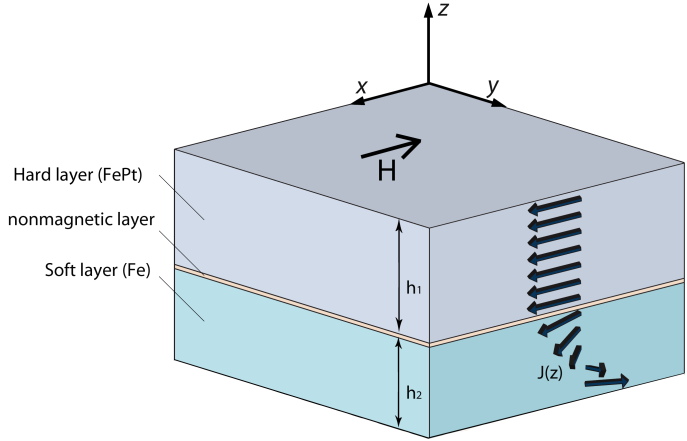

**Figure 1:** ECM multilayered system under investigation.

Defects observed in real films [36], [37] are mostly easy axis deviations. As it is seen from the [36] the full width at half maximum (FWHM) of the easy axis distribution function $f(\alpha)$ for the easy axis could be estimated as a value from $3\pi/10$ to $4\pi/10$. Therefore for the easy axis deviation the normal distribution was chosen with $\sigma = 0.9$ radian and maximum deviation angle equal to 3 radian.

We have also performed calculations in order to get optimum values for defects. Besides the easy axis defects which takes an absolutely major part in influence on structure properties there were two defects that were found to play a significant role: fluctuations of the saturation magnetization and anisotropy constant. We have considered exchange constant fluctuations as well although they were not found to change the parameters of structure in any way. Therefore we have performed the coercitivity field as a function of the saturation magnetization standard deviation and as a function of the anisotropy constant standard deviation and estimated the optimum values of defects for the model under investigation.

According to experiments [36] the biggest value of the FePt film coercive field with still mazelike or continuous morphology corresponds to the film thickness equal to 10 nm which we have chosen for our study. Thus considered FePt film has a granular continuous morphology. The FePt film became discontinuous and forms mazelike connected "nanoislands" with average size 60 nm for the material with the perpendicular anisotropy. For our structure where hard magnetic layer has planar anisotropy this value was taken to be 20 nm. An example of such a granular FePt film with defects added is presented on Fig. 2.

As a conclusion the following defects were chosen:
1. The easy axis deviation defect with normal distribution which has $\sigma = 0,9$ rad and

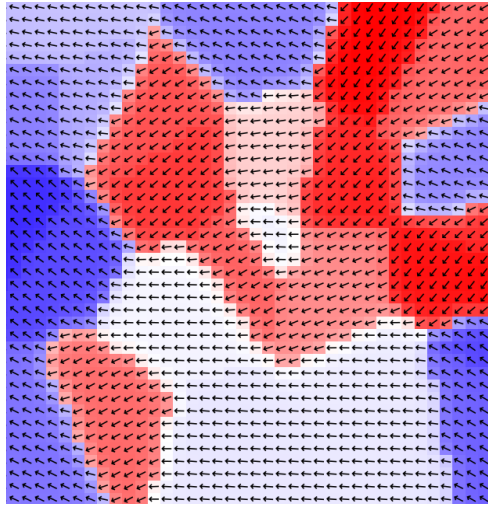

**Figure 2:** Granular FePt film consisted of nanoislands of a random shape which had average size of 20 nm. The defects were added separately to the every nanoisland. Arrows indicates the magnetization vector and the color indicates negative (red) and positive (blue) magnetization vector y-component. Easy axis is along x-axis.

3 rad as a maximum;
2. The saturation magnetization deviation defect with normal distribution which has $\sigma = 10 \ emu/cm^3$ and $100 \ emu/cm^3$ as a maximum;
3. The anisotropy constant deviation defect with normal distribution which has $\sigma = 3 \times 10^6 \ emu/cm^3$ and $6 \times 10^6 \ emu/cm^3$ as a maximum;

All the defects were added separately to the every nanoisland of a random shape which had average size of 20 nm. As far as defects affect mainly coercive field they were added to hard layer only.

In the paper [39] the phase of four different samples containing FePt were studied before and after annealing and X-ray $\theta/2\theta$ diffraction spectra were performed which can assure one that the $L1_0$ phase of FePt could form under aforementioned conditions.

## 3. Results

As far as there is much more experimental data available for single layer loops the latter were used to make the model optimization and to chose the correct values for the defects. On the Fig 3(a) The hysteresis loop for the FePt layer of nominal thickness with $t_{FePt} = 10 \ nm$ is shown and on the Fig 3(b) the same loop is shown after selected defects were added to the model. Although the coercitivity field has not changed dramatically being 5.6 T in the first model it became 5 T after model optimization, - the energy product has changed significantly. Before the model optimization the energy product $(BH)_{max}$ was $17,30 \times 10^7 \ erg/cm^3$, and after the optimization $(BH)_{max} = 6,14 \times 10^7 \ erg/cm^3$. So one can see that it has dropped almost three times and becomes closer to the value that could be expected from some experimental works, although done not on this particular structure ([15], [16], [36], [37]).

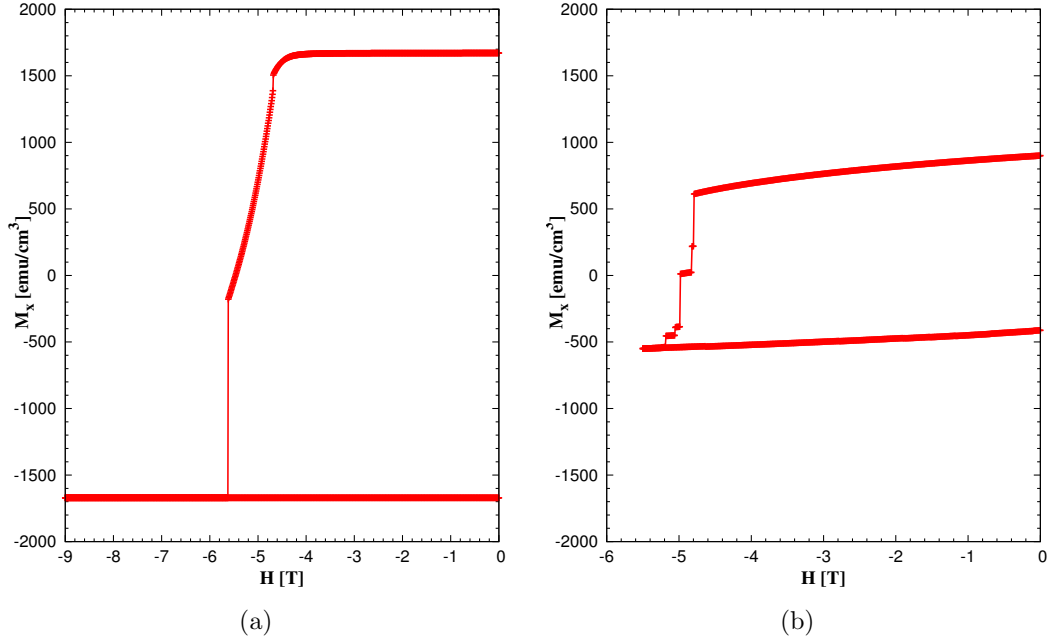

**Figure 3:** Calculated hysteresis half-loop (loops are symmetric) for $L1_0$-FePt layer with thickness $t_{FePt} = 10$ $nm$ at zero temperature. External field is applied in plane. 3D-micromagnetic model without defects(a) and with defects(b) are used.

On the Fig 4(a) the hysteresis loop for bilayer FePt/Fe with nominal thicknesses $t_{FePt} = 10$ $nm$ and $t_{Fe} = 2$ $nm$ is presented. It is clearly seen that the obtained hysteresis loop is quite different from those for single layers. The coercivity field $H_C$ is 3,5 T which is significantly smaller than that for FePt layer but much higher than with $H_C = 200$ Oe obtained for single Fe layer(graph not presented here). The saturation magnetization $M_S$ also shows a value between those for Fe and FePt single layers and is around 1250 emu/cm$^3$. The hysteresis loop has a rectangular shape which indicates that two magnetic layers have been exchange coupled and so called spin spring is suppressed which is very good for applications considered in this paper.

It is important for permanent magnet application to determine structure parameters which could provide the maximum energy product for the ECM. So for the FePt/Fe with $t_{FePt} = 10$ $nm$ and $t_{Fe} = 2$ $nm$ bilayer it is: $(BH)_{max} = 6,56 \times 10^7$ $erg/cm^3$ which is higher than for a single layer. It was important to study the structures with different layer thicknesses in order to choose the optimum values and to predict the behavior of the magnetic curves in case of slight change of thicknesses. So the structures FePt/Fe with nominal thicknesses 10 nm/4 nm and 10 nm/6 nm were also studied and there hysteresis loops are presented on Fig 4(b) and Fig 4(c) correspondingly.

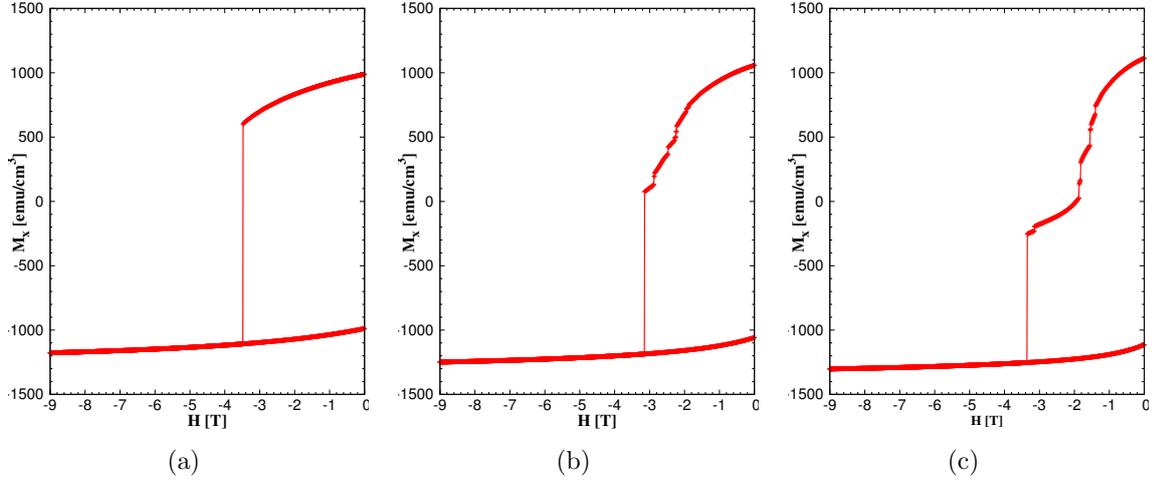

**Figure 4:** Calculated hysteresis half-loops for the $L1_0$-FePt/Fe bilayer with thicknesses (a)10 nm/2 nm, (b) 10 nm/4 nm and (c) 10 nm/6 nm at zero temperature. External field is applied in plane. 3D-micromagnetic model with defects added.

## 4. Discussions

Depending on geometrical parameters of ECM different regimes could be realized. The magnetization reversal process in ECM was studied in [40]. If the thickness of soft layer is small enough the formation of such spin springs would become energetically insufficient. This regime is called rigid magnet and is the regime of interest for the permanent magnets applications such as motors, etc.

The critical value for soft layer thickness $t_0$ was estimated in the work [41] to be close to the thickness of Bloch wall for the hard phase of ECM.

On the Fig 4(a) one can see the loop shape which corresponds to Rigid Magnet regime. In this case soft layer is thin enough for spin spring to be suppressed. Soft layer becomes magnetized collinearly to the hard layer and rotates only with the latter in external fields higher than the nucleation field. For this regime the values of the bias field $H_{ex}$ and the nucleation field $H_N$ coincide.

One can see on the Fig 4(c) that when soft layer thickness $t_{Fe}$ becomes 4 nm there are two significant steps of magnetization with the field, so there are two different values for the nucleation field and bias field and the hysteresis loop has a shape specific for the ECM with the soft layer thickness $t_{Fe} \geq t_0$. This indicates that exchange coupling of layers is not sufficient and soft layer rotates reversibly in external field higher than the bias field $H > H_{ex}$ and forms the spin spring. The part of loop which responses to the demagnetization would be reversible until external field is lower than the nucleation field $H > H_N$. For the bilayer with the soft layer thickness $t_{Fe} = 6\ nm$ shown on the Fig 4(c) described mechanism is much more evident.

Indeed, if one calculates energy products for these structures the loss of exchange coupling would become apparent. Thus, for FePt/Fe with Fe layer thickness $t_{Fe} = 4$ $nm$ $(BH)_{max} = 5,78 \times 10^7$ $erg/cm^3$ and for the FePt/Fe with $t_{Fe} = 6$ $nm$ $(BH)_{max} = 3,84 \times 10^7$ $erg/cm^3$, while for the FePt/Fe with $t_{Fe} = 2$ $nm$ this value was $(BH)_{max} = 6,56 \times 10^7$ $erg/cm^3$.

## 5. Summary

In summary, we have systematically studied the dependence of the magnetic properties for the exchange-coupled $L1_0$-FePt/Fe bilayer with planar anisotropy at zero temperatures on defects. The whole range of the most valuable defects for micromagnetic simulations such as the easy axis deviation and the saturation magnetization, the anisotropy constant and exchange constant fluctuations has been considered for different thicknesses of soft Fe layer. The dependence of the hysteresis shape on the soft layer thickness was studied and the spin spring was found to be suppressed and the rigid magnet regime to occur for thickness of the soft layer $t_{Fe} = 2$ $nm$. Also, a maximum coercivity and energy product are obtained for the structure with a soft layer film thickness $t_{Fe}$ equal to 2 $nm$.

## 6. ACKNOWLEDGMENTS

The work was supported by the Ministry of Education and Science of the Russian Federation (Grants nos. 11.519.11.1011, 11.519.11.3003)